\documentclass[conference]{IEEEtran}

\ifCLASSINFOpdf

\else

\usepackage{amssymb}
\usepackage{algorithm,algorithmic}
\usepackage{cite}
\usepackage[dvips]{graphicx}
\usepackage{amsmath}
\usepackage{multirow}
\usepackage{subfigure}
\usepackage{color}
\usepackage[left=0.71in,top=0.94in,right=0.71in,bottom=1.18in]{geometry}
\begin{document}

\title{Dynamic Sleep Control in Green Relay-Assisted Networks for Energy Saving and QoS Improving}

\author{\IEEEauthorblockN{Fang Chen, Bo Yang, Qiaoni Han, Cailian Chen and Xinping
Guan}\\
\vspace{-0.2cm}
\IEEEauthorblockA{Department of Automation, Shanghai Jiao Tong University, Shanghai, China\\Key Laboratory of System Control and Information Processing, Ministry of Education of China, Shanghai, China \\
Email: \{chenfang0007, bo.yang, qiaoni, cailianchen, xpguan\}@sjtu.edu.cn }}
\maketitle

\begin{abstract}
  We study the relay station (RS) sleep control mechanism targeting on reducing energy consumption while improving users' quality of service (QoS) in green relay-assisted cellular networks, where the base station (BS) is powered by grid power and the RSs are powered by renewable energy. By adopting green RSs, the grid power consumption of the BS is greatly reduced. But due to the uncertainty and stochastic characteristics of the renewable energy, power supply for RSs is not always sufficient. Thus the harvested energy needs to be scheduled appropriately to cater to the dynamic traffic so as to minimize the energy saving in the long term. An optimization problem is formulated to find the optimal sleep ratio of RSs to match the time variation of energy harvesting and traffic arrival. To fully use the renewable energy, green-RS-first principle is adopted in the user association process. The optimal RS sleeping policy is obtained through dynamic programming (DP) approach, which divides the original optimization problem into per-stage subproblems. A reduced DP algorithm and a greedy algorithm are further proposed to greatly reduce the computation complexity. By simulations, the reduced DP algorithm outperforms the greedy algorithm in achieving satisfactory energy saving and QoS performance.
 \end{abstract}

\IEEEpeerreviewmaketitle
\section{Introduction}
 With the rapid growth of energy demand for information and communication technology (ICT), energy-saving approaches for mitigating energy consumption are urgently required. In recent years, green communications have been proposed to improve the energy efficiency \cite{Raffaele2011Thepotential}. Renewable energy utilization has emerged as a promising candidate, which has attracted more and more attention from both academia and industry. By exploiting renewable energy, e.g., solar energy, wind energy and so on, traditional energy consumption will be significantly reduced. The energy harvesting technology, termed as ``energy harvesting'', is proved to be practical, i.e., the energy obtained from the surrounding environments can potentially support sustainable operations of wireless communication equipments or devices. From the measurement results of renewable energy sources, energy harvesting level can sustain the operation of a base station, especially for small cells  \cite{Giuseppe2013Hetnets}. Therefore, we introduce renewable energy to power the relay stations (RSs) to reduce traditional energy consumption. In case of the shortage of RSs' renewable energy, the BS is powered by grid power so as to guarantee the users' service.

 However, there are two key challenges need to be addressed for the utilization of renewable energy in cellular networks: firstly, renewable energy harvesting is highly dynamic and uncertain, which influences the transmission directly; secondly, the traffic load shares the characteristic of dynamic variation in both time and space domain. Thus there exists a mismatch between the traffic variation and energy dynamics. On the other hand, since the energy consumption of a station mainly comes from baseband signal processor, controller, air-conditioner and etc., rather than transmit power \cite{Holger2003Anoverview}, turning stations into sleep mode when the traffic load is low, is considered as an effective way to save the energy consumption. Then, for green RSs equipped with batteries, the saved energy in sleep mode will be stored for future use when the traffic load is high so as to save more energy in the long term.

 Compared with the BS that is supplied by grid power, RSs cover a much smaller area, require lower transmit power and have no wired backhaul connection \cite{Roberto2011Energy}. Moreover, users served by RSs mostly experience much higher average signal-to-interference-plus-noise-radios (SINRs) \cite{Mikio2010Relay}. All these good features of RS make it a good option to save energy consumption in cellular networks while improving users' QoS, especially for cell-edge users. The issue of green RSs has been discussed in different aspects. The green RS selection was introduced in cooperative communication networks in \cite{Bhargav2010Voluntary}. A deterministic energy harvesting model for the Gaussian RS channel was considered in \cite{Chuan2013Throughput},\cite{Chuan2012Outage}
 where delay and no-delay constrained traffic were studied. In \cite{Kaya2013Cooperative}, cooperative communication with energy harvesting nodes using an energy sharing strategy was studied. The concept of energy transfer in green RS systems was considered in \cite{Berk2013Energy}, where an offline power allocation scheme was proposed. But the sleep control for green RSs aiming at both energy saving and QoS improving has not been much studied yet.

 In this paper, we consider a network scenario to reduce the grid power consumption of cellular networks and improve users' QoS at the same time. By introducing renewable energy powered RSs, the grid power consumption of the BS is greatly reduced. And by deploying RSs at the edge of the BS's coverage where the channel conditions between users and the BS are usually poor, users' QoS can be significantly improved with the assistance of the RSs. The main contributions of this paper can be summarized as follows.
  \begin{itemize}
  \item We formulate a weighted grid power consumption and blocking probability minimization problem by taking into account the users' QoS in terms of blocking probability.
  \item A dynamic RS sleep control mechanism based on DP approach is proposed, and a reduced DP algorithm as well as a greedy algorithm is further provided to greatly reduce the computation complexity.
 \end{itemize}

 The remainder of this paper is organized as follows. In Section \ref{sec:model}, the system model is described. In section \ref{sec:formulation}, we formulate the optimization problem. Section \ref{sec:algorithm} presents a DP algorithm, and a reduced DP algorithm as well as a greedy algorithm. The simulation results are given in Section \ref{sec:simulation} and the conclusion is drawn in Section \ref{sec:conclusion}.

\begin{figure}
    \centering
        \includegraphics[width=0.25\textwidth]{{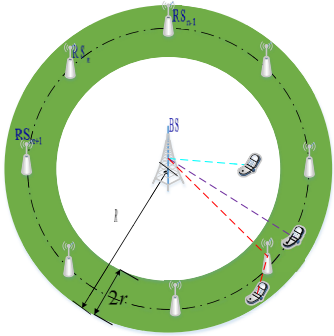}}
        \vspace{-0.4cm}
        \caption{system model
        \label{fig:system}}
        \vspace{-0.7cm}
\end{figure}

\section{System Model}\label{sec:model}
 We consider a downlink wireless system, which consists of a BS powered by grid power, $N$ green RSs powered by renewable energy. Note that this paper mainly considers the QoS improvement of cell-edge users, whose channel conditions to the BS are usually poor. Therefore, RSs are deployed at the edge of the BS's coverage area and assumed to be uniformly distributed as depicted in Fig.\ref{fig:system}. The distances between RSs and the BS are assumed to be the same, denoted by $(R-2r)$, where $R$ denotes the coverage radius of the BS and $2r$ denotes the width of RSs' covering area. Let $\mathcal{N}=\{1,\cdots,N\}$ denote the set of RSs, and $\mathcal{M}=\{1,\cdots,M\}$ denote the set of users. The coverage area of the BS and the $n$-th RS are denoted as $\mathcal{A}_0$ and $\mathcal{A}_n$ respectively, and the area with no RS covering in $\mathcal{A}_0$ as $\widetilde{\mathcal{A}}_0$. Users in $\widetilde{\mathcal{A}}_0$ can only be served by the BS while users in $\mathcal{A}_n$ can be served by either the $n$-th RS or the BS. Each RS has two state modes (active or sleep), and a RS in active mode can turn to opportunistic sleep mode. The operational time line (e.g., a period of 24 hours) is divided into $I$ time slots (each with index $i$ and the length $L^{(i)}$).


\subsection{Traffic and Channel Model}
 In each region, users are considered uniformly distributed and arrive randomly according to a Poisson distribution with a certain arrival rate. The arrival rate in area $\widetilde{\mathcal{A}}_0$ and $\mathcal{A}_n$ at time $t$ are denoted as $\lambda_0(t)$ and $\lambda_n(t)$ respectively.
 Each user has a minimum rate requirement $r_0$, and the transmission duration of each user follows exponential distribution with mean $1/\mu$. Thus the arriving traffic load in $\widetilde{\mathcal{A}}_0$ and $\mathcal{A}_n$ are calculated as $\lambda_0/\mu$ and $\lambda_n/\mu$. All users are assumed to remain stationary until the transmission is completed.

 For the BS, it serves a user through the link between the BS and the user, which we term it as the direct link (DL). For the RSs, each RS accesses the BS through a wireless backhaul link (BL) and forwards data to its serving users through a wireless access link (AL). The spectrum efficiency (transmission rate per unit bandwidth) for user $m$ on DL is expressed as
 \begin{equation}\label{equ:spectrum efficiency on DL}
  \small{C_{0m}^{DL} = \log_2(1+\frac{P^t_0\beta_1l_{0m}^{-\alpha_1}}{\sigma^2+I_{0m}})}
 \end{equation}
 where $P^t_0$ is the transmit power of the BS; $\beta_1$ and $\alpha_1$ are the path-loss constant and path-loss exponent on DL respectively; $\sigma^2$ is the noise power; and $I_{0m}$ is the interference power. We assume that the interference is well taken care of by certain interference management tool and is randomized as noise.

 Similarly, the spectrum efficiency on AL and BL are
 \begin{equation}\label{equ:spectrum efficiency on AL}
  \small{C_{nm}^{AL} = \log_2(1+\frac{P^t_n\beta_2l_{nm}^{-\alpha_2}}{\sigma^2+I_{nm}})}
 \end{equation}
 \begin{equation}\label{equ:spectrum efficiency on BL}
  \small{C_{0n}^{BL} = \log_2(1+\frac{P^t_0\beta_3l_{0n}^{-\alpha_3}}{\sigma^2+I_{0n}})}
 \end{equation}

 As the long time-scale performance is considered here, we ignore the fast fading effects and assume that the spectrum efficiency is constant during the transmission \cite{Jie2012Adynamic}. Moreover, the transmit power is considered to be fixed. Based on the above assumptions, the spectrum efficiency ($C_{0m}^{DL}$, $C_{nm}^{AL}$, $C_{0n}^{BL}$) only depends on the link distance, which is independent of the time period.

 Considering the minimal data rate requirement of each user, the corresponding bandwidth demand imposed on different access links are calculated as
 \begin{equation}\label{equ:bandwidth demand}
  \omega_{0m}^{DL} = \frac{r_0}{C_{0m}^{DL}},~~
  \omega_{nm}^{AL} = \frac{r_0}{C_{nm}^{AL}},~~
  \omega_{0n}^{BL} = \frac{r_0}{C_{0n}^{BL}}
 \end{equation}

 We assume that the BS and each RS have limited radio resource, denoted as $\small{W^{th}_0}$ and $\small{W^{th}_n}$ correspondingly. If the total bandwidth demand of the users served by the station exceeds its resource limit, some of its users will be blocked.

 \subsection{User Association and RS State Model}
 To fully utilize RSs' renewable energy and save the BS's grid power, RS-first principle is adopted in the user association process, that is, all users within the coverage of a RS associate the RS first, and only when the RS is in sleep mode do they switch to the BS.

 For the green RSs, since the energy harvesting and the traffic arrival are both dynamic. There may exist a mismatch between them. Thus, RSs can turn to sleep mode with a sleep probability denoted as $\boldsymbol{\varphi}^{(i)}$ in the $i$-th time slot, and it can be implemented in time domain. That is, if the traffic load is low and the renewable energy is insufficient, an active RS will turn to sleep mode for a fraction of time $\varphi^{(i)}_n$ in the time slot to save energy, and the saved energy can be stored in the battery for future use. The capacity of batteries equipped with RSs are considered to be the same and denoted by $B^{max}$. Thus, the energy sources of a RS are the battery and the energy harvesting at that time slot.

 When a RS turns to sleep mode, users within its range will be served by the BS. The overview of the system operation is as follows: The BS is active in all time slots, while the RSs' sleep probability $\boldsymbol{\varphi^{(i)}}$ is decided at the beginning of time slot $i$. Moreover, according to the RSs' state mode, user association process is performed as: users arrive in $\widetilde{\mathcal{A}}_0$ during the time slot will associate with the BS directly and those in $\mathcal{A}_n$ associate with the corresponding RS unless it is in sleep mode.

 \subsection{Energy Consumption Model}
 We adopt the model in EARTH project \cite{Muhammad2012Energy}, the power consumption model of a station consists of two categories: the constant part and the dynamic part relevant to the traffic load. Thus, the energy consumption of the BS and the $n$-thRS are respectively expressed as
 \begin{equation}\label{equ:energy consumption}
  \small{P_0 = P_{0,sta}+\triangle_bP_{0,tra}},~~
  \small{P_n = P_{n,sta}+\triangle_rP_{n,tra}}
 \end{equation}
 where $P_{0,sta}$, $P_{n,sta}$ are the fixed part when the station is in active mode; $\triangle_bP_{0,tra}$, $\triangle_rP_{n,tra}$ are the variable part related to traffic load; and $\triangle_b$, $\triangle_r$ are the variable energy consumption slope. The typical values of $\triangle_b$ and $\triangle_r$ are different. Considering fixed transmission power of both BS and RS, the dynamic power consumption $P^{(i)}_{0,tra}$ and $P^{(i)}_{n,tra}$ are proportional to the bandwidth utility $W^{(i)}_0/W^{th(i)}_0$ and $W^{(i)}_n/W^{th(i)}_n$, which are expressed as
 \begin{equation}\label{equ:dynamic power consumption}
 \small{P^{(i)}_{0,tra} = P_0^tW^{(i)}_0/W^{th(i)}_0},~~
 \small{P^{(i)}_{n,tra} = P_n^tW^{(i)}_n/W^{th(i)}_n}
 \end{equation}
 where $W_0^{(i)}$, $W_n^{(i)}$ are the resource utilization; $W_0^{th(i)}$, $W_n^{th(i)}$ are the resource limitation (resource allocated to the station). As the total bandwidth is shared by the BS and RSs, resource allocated to each station is proportional to its arriving traffic load.

 The power consumption of a RS in sleep mode is assumed to be constant, denoted as $P_s$. Then the total energy consumption of the BS and RS$n$ in time slot $i$ are expressed as
 \begin{equation}\label{equ:energy consumption of BS}
  \small{E^{(i)}_0 = (P_{0,sta}+\triangle_bP_0^tW^{(i)}_0/W^{th(i)}_0)L^{(i)}}
 \end{equation}
 \begin{equation}\label{equ:energy consumption of RS}
  \small{E^{(i)}_n\!=\![(P_{n,sta}\!+\!\triangle_rP_n^tW^{(i)}_n/W^{th(i)}_n)(1\!-\!\varphi^{(i)}_n)\!+\!P_s\varphi^{(i)}_n]L^{(i)}}
 \end{equation}
 where the expressions of $W^{(i)}_0$, $W^{(i)}_n$ can be referred to in the Appendix. The detailed derivation is referred to in \cite{Jie2012Adynamic}, and brief supplementary explanation is given in the Appendix.

 \subsection{Blocking Probability Analysis}
 The blocking probability is defined as the probability that a newly arrived user is blocked, i.e., none of the active stations (BS and RSs) can provide service to the user. For simplicity we ignore the time index in this section.

 In area $\widetilde{\mathcal{A}}_0$, with the proposed user association scheme, a newly arrived user will be blocked if the required bandwidth exceeds the bandwidth limit of the BS. Therefore, the blocking probability of users in area $\widetilde{\mathcal{A}}_0$ in time slot $i$ is
 \begin{equation}\label{equ:bolck probability}
 \small{P_{0,blk} = Pr(w_{0m'}+\sum \limits_m a_{0m}w_{0m} \geq W^{th}_0)}
 \end{equation}
 where $w_{0m'}$ is the bandwidth demand of a new user $m'$ and $a_{0m}$ is the binary association variable which equals 1 if user $m$ is associated to the BS and 0 otherwise.

 The expression of $P_{0,blk}$ is derived as below
 \begin{equation}\label{equ:blocking probability of BS}
 \small{P_{0,blk} = (\frac{\lambda'_0}{\lambda'_0+\mu})^{\lceil W^{th}_0/\gamma_0\rceil}}
 \end{equation}
 where $\footnotesize{\lambda'_0}$, $\gamma_0$ is referred to in \cite{Jie2012Adynamic} and the Appendix.


  For renewable energy powered RSs, a blocking event can be caused by either resource limitation or energy depletion. The blocking probability of an active RS caused by resource limitation $\widehat{P}_{n,blk}$ can be calculated similarly as $P_{0,blk}$, which is expressed as
  \begin{equation}\label{equ:resource blocking probability of RS}
  \small{\widehat{P}_{n,blk} = (\frac{\lambda_n}{\lambda_n+\mu})^{\lceil W^{th}_n/\gamma_n\rceil}}
  \end{equation}
  where $\gamma_n$ can be found in the Appendix.

  If the energy harvesting is not sufficient, the RS will turn to sleep with the  probability $\varphi_n$. Denoting the energy harvesting rate and the energy consumed from the battery as $H_n$ and $C_n$ respectively, the maximal available energy of a RS is $C_n+H_nL=(1-\varphi_n)P_nL+\varphi_nP_sL$ \cite{Jie2014Base}. Thus, the maximal sleep ratio can be expressed as
  \begin{equation}\label{equ:sleep ratio}
   \small{\varphi_n = \max\{0,\frac{P_n-(C_n/L+H_n)}{P_n-P_S}\}}
  \end{equation}

  Hence, the blocking probability for a green RS is
  \begin{equation}\label{equ:blocking probability of RS}
  P_{n,blk}= \varphi_n+(1-\varphi_n)\widehat{P}_{n,blk}
  \end{equation}

  As the network is divided into multiple areas and the users are uniformly distributed, the system blocking probability can be calculated as
  \begin{equation}\label{equ:system blocking probability}
  \small{P_{blk} = P_{0,blk}Pr(m'\in\widetilde{\mathcal{A}}_0)+\sum\limits_{n=1}^NP_{n,blk}Pr(m'\in\mathcal{A}_n)}
  \end{equation}
  where $Pr(m'\in\widetilde{\mathcal{A}}_0)$ and $Pr(m'\in\mathcal{A}_n)$ are the probability a newly arrived user belongs to area $\widetilde{\mathcal{A}}_0$ and $\mathcal{A}_n$ respectively. As the users are assumed uniformly distributed and arrive according to a Poisson distribution, the probability can be expressed as $Pr(m'\in\widetilde{\mathcal{A}}_0)=\lambda_0/(\lambda_0+\sum\limits_{n=1}^N\lambda_n)$, $Pr(m'\in\mathcal{A}_n)=\lambda_n/(\lambda_0+\sum\limits_{n=1}^N\lambda_n)$.

 \section{Problem Formulation}\label{sec:formulation}
 Our objective is to minimize the grid power consumption and system blocking probability with the resource constraints of stations (BS and RSs) and renewable energy constrains of RSs.
 Then, the optimization problem can be considered as follows: Given the traffic profile $\small{\boldsymbol{\lambda}_0 = \{\lambda_0^{(i)}\}^{i=1,\cdots,I}}$, $\small{\boldsymbol{\lambda} = \{\lambda_n^{(i)}\}^{i=1,\cdots,I}_{n=1,\cdots,N}}$ and the renewable energy harvesting profile $\small{\boldsymbol{H}_n = \{H_n^{(i)}\}^{i=1,\cdots,I}_{n=1,\cdots,N}}$, we aim to minimize the total grid power consumption (i.e. the power consumption of the BS) and users' blocking probability, through deciding the RSs' sleep ratio $\small{\boldsymbol{\varphi} = \{\varphi_n^{(i)}\}^{i=1,\cdots,I}_{n=1,\cdots,N}}$. Therefore, the optimization problem is formulated as
 \begin{align}
  \min\limits_{\boldsymbol{\varphi}} \quad & \sum_{i=1}^IE_0^{(i)}+\psi\sum_{i=1}^I\omega^{(i)}P_{blk}^{(i)}\label{equ:original}
 \end{align}
 where $\psi$ is the weighting parameter balancing the two objectives; and $\omega^{(i)}$ is the weighting factor reflecting the system sensitivity to blocking probability in each time slot, which satisfies $\sum_{i=1}^I\omega^{(i)}=1$.

  By adjusting $\psi$, the relative importance of the blocking probability minimization is balanced with the grid power reduction. And by considering weighted blocking probability, the QoS can be adjusted more flexibly which can satisfy different QoS requirements at different time periods.

 \section{RS Sleep Algorithm}\label{sec:algorithm}
 The problem is to optimize the total energy consumption and blocking probability in a long term by deciding the blocking probability of each time slot, which is complicated to slove. As dynamic programming approach has advantages in dividing the whole problem into simple per-stage sub-problems \cite{Dimitri2005Dynamic}, we propose the sleep algorithm based on the DP approach to find the optimal policy of the problem.

 \subsection{Dynamic Programming Algorithm}
 The DP algorithm contains three key components: state, action and cost function. In our problem, the state is the battery state (energy level), denoted as $\boldsymbol{B}$. The action is the adjustment of sleep ratio $\boldsymbol{\varphi}$. The per-stage cost is the weighted combination of grid power consumption and system blocking probability, which is expressed as
 \begin{equation}\label{equ:cost function of DP}
  \small{c^{(i)}(\boldsymbol{B}^{(i)},\boldsymbol{\varphi}^{(i)}) = E_0^{(i)}+\psi\omega^{(i)}P_{blk}^{(i)}}
 \end{equation}

 As we described above, the DP algorithm breaks the original problem down into sub-problems with respect to the stage (i.e., time slot in our problem). Then we perform a backward induction of the cost-to-go functions from time slot $T$ to 1, where the objective of each time slot is to minimize the cost of the current time slot and that of the following slots. Thus the cost-to-go function is defined recursively as
 \begin{equation}\label{equ:cost-to-go function of DP}
 \small{J^{(i)}(\boldsymbol{B}^{(i)})\!\!=\!\!\begin{cases}\min\limits_{\boldsymbol{\varphi}^{(i)}}c^{(i)}(\boldsymbol{B}^{(i)},\boldsymbol{\varphi}^{(i)}),\!&\!\!{\!\!i\!=\!\!I}\\ \min\limits_{\boldsymbol{\varphi}^{(i)}}\{c^{(i)}(\boldsymbol{B}^{(i)},\boldsymbol{\varphi}^{(i)})\!\!+\!\!J^{(i+1)}(\boldsymbol{B}^{(i+1)})\},\!\!&\!\!{\!i\!\!<\!\!I}
   \end{cases}}
 \end{equation}
 where $\footnotesize{J^{(i)}(\boldsymbol{B}^{(i)})}$ denotes the minimal cost for the subproblem with $\small{\boldsymbol{B}^{(i)}}$ as its initial state.

 Considering RS's battery capacity $B^{max}$, the saved energy in sleep mode can be stored in the battery for future use. Thus, the current stage's state is a function of the last stage's state and action, expressed as $B_n^{(i+1)}=f(B_n^{(i)},\varphi^{(i)}) = \min\{B^{max}, B_n^{(i)}+H_n^{(i)}L^{(i)}- [(1- \varphi_n^{(i)})P_n^{(i)}L^{(i)}+\varphi_n^{(i)}P_sL^{(i)}]\}$.
  Then, by conducting backward induction of Eq.(\ref{equ:cost-to-go function of DP}) from time slot $I$ to 1, we can obtain the minimum cost equal to $\footnotesize{J^{(1)}(B^{(1)})}$, which is the optimal solution of the original optimization problem.

  Due to the difficulty of solving this non-convex problem, the continuous value of sleep ratio is hard to obtain. As the sleep ratio can be calculated by Eq.(\ref{equ:sleep ratio}), we discretize the energy consumed from the battery $C_n$ and denote the energy consumption unit as $C_{n0}$ which is set to be small. The range of $\footnotesize{C_n}$ is from $\footnotesize{-(H_n-P_s)}$ which means the sleep ratio equals to 1 and the harvested energy is stored in the battery, to $\footnotesize{\min\{B^{max},D_n^{max}-H_n\}}$ where $\footnotesize{D_n^{max}}$ is the maximal demanding energy when the sleep ratio equals to 0. Therefore, the candidate actions in slot $i$ is $\footnotesize{K_n=\min\{B^{max},D_n^{max}-H_n\}/C_{n0}+(H_n-P_s)/C_{n0}}$. Thus, the action and state space of the DP algorithm in each stage are both $\footnotesize{\prod\limits_{n=1}^NK_n^{(i)}}$. The computational requirement is still overwhelming especially when the number of RSs is large. Thus it is difficult to implement the standard DP algorithm to obtain the optimal solution. Then, in the following section we introduce the reduced DP algorithm to simplify the decision process.

 \subsection{Reduced DP Algorithms}
 As the users will be switched only to the BS when a RS turn to sleep mode, the sleep of a RS has no influence on other RSs. Therefore, we develop the reduced DP algorithm which decides the per-RS sleep ratio iteratively.
 The per-RS cost function in time slot $i$ is as follows
 \begin{equation}\label{equ:cost function per-RS}
  \small{\tilde{c}_n^{(i)}(B_n^{(i)},\varphi_n^{(i)}) = \tilde{E}_{0n}^{(i)}+\psi\omega^{(i)}\tilde{P}_{blk,n}^{(i)}}
 \end{equation}
 where $\tilde{E}_{0n}$, $\tilde{P}_{blk,n}$ is corresponding to only one RS and the expression is omitted here which can be obtained through replacing the summation of all RSs in Eq.(\ref{equ:energy consumption of BS}) and Eq.(\ref{equ:system blocking probability}) by one RS.
 The per-RS's cost-to-go function is
 \begin{equation}\label{equ:cost-to-go function per-RS}
  \small{\tilde{J_n}^{(i)}(B_n^{(i)})\!\!=\!\! \begin{cases}\min\limits_{\varphi_n^{(i)}}\tilde{c}_n^{(i)}(B_n^{(i)},\varphi_n^{(i)}),\!&\!\!\!\!{i\!\!=\!\!I}\\ \min\limits_{\varphi_n^{(i)}}\{\tilde{c}^{(i)}(B_n^{(i)},\varphi_n^{(i)})\!\!+\!\!\tilde{J}_n^{(i+1)}(B_n^{(i+1)})\},\!&\!\!\!\!{i\!\!<\!\!I}
   \end{cases}}
 \end{equation}

 The cost-to-go function of Eq.(\ref{equ:cost-to-go function of DP}) is approximated as the summation of per-RS's cost-to-to function, i.e.,
 \begin{equation}\label{equ:approximation}
  \small{J^{(i)}(\boldsymbol{B}^{(i)}) \approx  \tilde{J}^{(i)}(\boldsymbol{B}^{(i)})}
 \end{equation}
 \begin{equation}\label{equ:cost-to-go function of reduced DP}
  \small{\tilde{J}^{(i)}(\boldsymbol{B}^{(i)})\!=\!\begin{cases}\min\limits_{\boldsymbol{\varphi}^{(i)}}\sum\limits_{n=1}^N\tilde{c}_n^{(i)}(B_n^{(i)},\varphi_n^{(i)}),\!\!\!&\!\!{i=I}\\ \min\limits_{\boldsymbol{\varphi}^{(i)}}\sum\limits_{n=1}^N\{\tilde{c}_n^{(i)}(B_n^{(i)},\varphi_n^{(i)})\!+\!J^{(i+1)}(B_n^{(i)},\varphi_n^{(i)})\},\!\!\!&\!\!{i<I}
   \end{cases}}
 \end{equation}

 The basic idea of the reduced DP algorithm is to find the optimal local action of each RS iteratively. The detailed description of the algorithm is summarized in Algorithm~\ref{alg:DP}. It reduces the action space from $\footnotesize{\prod\limits_{n=1}^NN_{c,n}^{(i)}}$ to $\footnotesize{\sum\limits_{n=1}^NN_{c,n}^{(i)}}$.
 \vspace{-0.3cm}
 \begin{algorithm}[htb]
 \caption{Reduced DP Algorithm}
 \textbf{Input}:{ $\boldsymbol{H}, \boldsymbol{\lambda}, \boldsymbol{B}^{(0)}$}\\
 \textbf{Output}:{ $\boldsymbol{\varphi}$}
 \begin{algorithmic}[1]
 \label{alg:DP}
 \FOR {$i=I$ to 1}
    \STATE Set $W_0^{th(i)}$, $W_n^{th(i)}$ according to the corresponding traffic load.
    \FOR {$n=1$ to $N$}
        \STATE Find the optimal local action of each RS.
            \IF {$i=I$}
                \STATE
                $\min\tilde{c}_n^{(i)}(B_n^{(i)},\varphi_n^{(i)})$ $\rightarrow \varphi_n^{(i)*}$
            \ELSE
                \STATE $\min\{\tilde{c}^{(i)}(B_n^{(i)},\varphi_n^{(i)})+\tilde{J}_n^{(i+1)}(B_n^{(i+1)})\}$ by search the candidate action space (adjust $C_n$ from $-(H_n-P_s)$ to $\footnotesize{\min\{B^{max},D_n^{max}-H_n\}}$)
                $\rightarrow \varphi_n^{(i)*}$
            \ENDIF
        \STATE Store the states and the corresponding cost-to-go function value.
    \ENDFOR
    \STATE Update states and go to next stage.
 \ENDFOR
 \end{algorithmic}
 \end{algorithm}
 \vspace{-0.3cm}

 By performing Algorithm~\ref{alg:DP}, the sleep ratio of each time slot is decided. Thus, when the RS sleep mechanism is executed from slot 1 to $I$, the corresponding sleep ratio of each RS, which is decided by Algorithm~\ref{alg:DP}, is selected at the beginning of each time slot.

 \subsection{Greedy Algorithms}
 The execution of the greedy algorithm performs from time slot 1 to $T$ that in each time slot it selects a action which minimizes the cost of the current stage without considering the following stages, i.e., every RS selects $\varphi_n^{(i)*}$ by minimizing the cost function of Eq.(\ref{equ:cost function per-RS}). Then it updates the state of next stage based on the selected action.

 As the action decision only considers the current stage, the solution might be suboptimal. But it is easy to perform and is worth considering if the requirement of algorithm accuracy is not too strict.

 \section{Simulation Results}\label{sec:simulation}
\begin{figure}
    \centering
        \includegraphics[width=0.4\textwidth]{{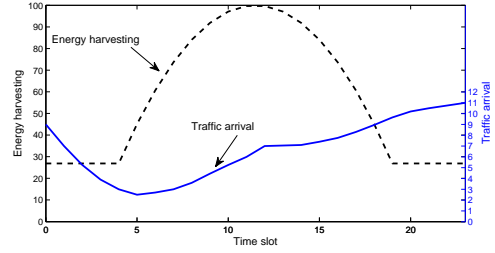}}
        \vspace{-0.3cm}
        \caption{Traffic arrival and Energy harvesting
        \label{fig:TrafficEnergy}}
        \vspace{-0.4cm}
 \end{figure}

 We consider a cellular network depicted as Fig.\ref{fig:system} where $R=800m$ and $r=100m$. The power consumption model is adopted from the EARTH project \cite{Muhammad2012Energy} and the channel model from LTE standard \cite{sesia2009lte}. For the BS, we set $P_{0,sta}=750w, \Delta_b=19.3, P_0^t=40W$. For the RSs, $P_{n,sta}=40w, \Delta_r=9.6, P_0^t=40W, P_s=10W$. The path-loss in AL, BL and DL are $PL^{dB}_{AL}=76.8+7.4log10(d_{nm})$, $PL^{dB}_{BL}=88.3+3.1log10(d_{nm})$ and $PL^{dB}_{DL}=91.3+3.4log10(d_{0m})$ respectively. The noise power density is considered -64.5 dBm/Hz, $r_0=200Kbps$ and $\mu=1s^{-1}$. The energy harvesting and traffic arrival follows the profile as illustrated in Fig.\ref{fig:TrafficEnergy} with different proportions. For simplicity, the proportion of all RSs are considered the same. The total bandwidth shared by the BS and RSs is 30MHz. And the resource allocation performed at the beginning of each time slot is according to their arriving traffic load.

\begin{figure}
    \centering
        \includegraphics[width=0.5\textwidth]{{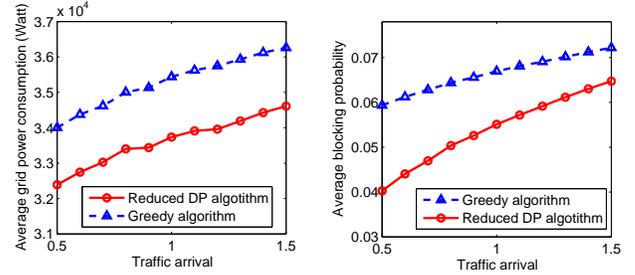}}
        \vspace{-0.8cm}
        \caption{Averagy grid power consumption and blocking probability versus time with different traffic arrival
        \label{fig:TrafficVariable}}
        \vspace{-0.4cm}
 \end{figure}

 To study the performance of the reduced DP algorithm and the greedy algorithm, the average grid power consumption and blocking probability are obtained with different traffic arrival rate which is depicted in Fig.\ref{fig:TrafficVariable}. It demonstrates that the average grid power consumption and the blocking probability both increase with the traffic arrival. Comparing the two algorithms, the reduced DP algorithm obviously achieves better performance with lower grid power consumption and blocking probability in various traffic arrival conditions. And the average grid power consumption gap between these two algorithms is not obvious in different traffic conditions while the average blocking probability gap decreases with the increase of the traffic arrival.

\begin{figure}
    \centering
        \includegraphics[width=0.4\textwidth]{{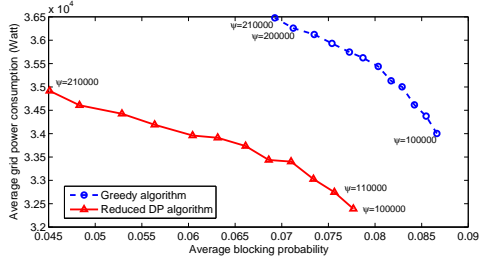}}
        \vspace{-0.3cm}
        \caption{Tradeoff curves between average grid power consumption and blocking probability
        \label{fig:PhiVariable}}
        \vspace{-0.6cm}
 \end{figure}

 As the optimization problem weights the power consumption and blocking probability, we obtain the tradeoff results as shown in Fig.\ref{fig:PhiVariable} by adjusting the weighting parameter $\psi$ ($\omega^{(i)}=1/I$). The results show that the more strict the blocking probability requirement is, the more grid power consumes. The reason is that when the blocking probability is loose, the system will sacrifice some QoS to minimize the grid power consumption. Comparing the two algorithms, the reduced DP algorithm outperforms the greedy algorithm which achieves both less grid power consumption and lower blocking probability in the same condition (i.e., the same weighting parameter). With the increase of the weighting parameter $\psi$, the blocking probability decreases while the grid power consumption increases.

\section{Conclusion}\label{sec:conclusion}
 We consider the dynamic RS sleep control mechanism aiming at reducing the grid power consumption while minimizing system blocking probability for a long term. The RSs' sleep ratio is appropriately scheduled via DP algorithm to cater to the energy harvesting and traffic arrival dynamic. A reduced DP algorithm as well as a greedy algorithm is further proposed to reduce the complexity. The reduced DP algorithm is shown to achieve better performance than the greedy algorithm. To further improve the performance of the reduced DP algorithm, joint optimization of all RSs' sleep control in the per-stage subproblems should be considered in the future. On the other hand.
 online dynamic RS sleep control without the statistical information of the energy harvesting and traffic arrival is included in the future work.

  \vspace{-0.1cm}
 \begin{appendix}
 \centerline{Blocking Probability Derivation}
 The derivation of blocking probability and the bandwidth utilization is referred to \cite{Jie2012Adynamic}. But the difference is that, in our network, whether the RS is active or not, bandwidth is required for the BS (through BL in active mode and DL otherwise). Thus, the bandwidth utilization of the BS consists of three parts: on DL for serving users in area $\widetilde{\mathcal{A}}_0$; on BL for serving users in $\mathcal{A}_n$ when RS$n$ is in active mode; on DL for serving users in $\mathcal{A}_n$ when RS$n$ is in sleep mode. We ignore the time slot index $i$ for simplicity. Then the expression of $W_0$ which is consisted of the three parts is derived as

 \begin{equation}\label{equ:W0}
 \small{\begin{array}{l}
  W_0\!=\!\int_0^{R - 2r} {\frac{{{r_0}}}{{{r_{0m}}}}} \frac{{{K_0}}}{{\pi {{(R - 2r)}^2}}}2\pi l{\rm{d}}l\!+\!\sum\limits_{n = 1}^N {\frac{1}{{{J_n}}}} \frac{{{r_0}{K_n}}}{{{r_{0n}}}}(1\!-\!{\varphi _n})\\~~~~~~~\!+\!\sum\limits_{n = 1}^N {\frac{1}{N}\int_{R - 2r}^R {\frac{{{r_0}}}{{{r_{0m}}(l)}}} } \frac{{{K_n}}}{{\pi {r^2}}}2\pi l{\rm{d}}l \varphi_n = K_0'\gamma_0
 \end{array}}
 \end{equation}
 where $\tiny{K_0}$, $\tiny{K_n}$ are the number of arriving users, which satisfies $\tiny{K_0/\lambda_0=K_n/\lambda_n}$ \cite{Jie2012Adynamic}; $K_0'$ is the total number of serving users; $\gamma_0$ means the average resource demand per user.

 Note that the coefficient $J_n$ is the multiple of resource demand for the BS when the users in area $\mathcal{A}_n$ is served by the BS directly (RS$n$ is in sleep mode) and when they are served by RS$n$ (RS$n$ is in active mode), i.e.,
 \begin{equation}\label{equ:Jn}
 \small{\int_{R-2r}^R\frac{r_0}{r_{0m}}\frac{K_n}{\pi r^2}2\pi l\text dl = J_n\int_{R-2r}^R\frac{r_0}{r_{0n}}\frac{K_n}{\pi r^2}2\pi l\text dl}
 \end{equation}
 where the left side is the resource demand for the BS on DL when RS$n$ is in sleep mode and its users are served by the BS, while the right side is that on BL when RS$n$ is in active mode and users are served by RS$n$. Thus when a RS turn to sleep mode from active mode, the resource demand for the BS multiplies $J_n$ times.

 As resource demand for the BS always exists whether users are served by the BS or RS, the actual total number of users served by the BS and the traffic load of the BS is
 \begin{equation}\label{equ:K0'}
 \small{K_0'=K_0+\sum\limits_{n=1}^N\frac{1}{J_n}K_n(1-\varphi_n)+\sum\limits_{n=1}^NK_n\varphi_n}
 \end{equation}
 \begin{equation}\label{equ:lambda_0'}
  \small{\lambda'_0 = \lambda_0+\sum\limits^N_{n=1} \frac{1}{J_n}\lambda_n(1-\varphi_n)+\sum\limits^N_{n=1} \lambda_n\varphi_n}
 \end{equation}

 Thus $\gamma_0$ can be calculated as
 \begin{equation}\label{equ:gamma_0}
 \small{
 \begin{split}
 &\gamma_0 =W_0/K_0' \\
 &\!=\!\frac{\frac{2r_0}{(R-2r)^2}\!\int^{R-2r}_0\!\! \frac{ldl}{r_{0m}(l)}\!+\!\frac{r_0}{\lambda_0}\!\!\sum\limits_{n=1}^N\!\frac{\lambda_n(1-\varphi_n)}{J_nr_{0n}}\!+\!\frac{2r_0}{Nr^2\lambda_0}\!\!\sum\limits_{n=1}^N\!\lambda_n\int_{R-2r}^R\frac{ldl\varphi_n}{r_{0m}(l)}}{1+\sum\limits_{n=1}^N\!\frac{\lambda_n(1-\varphi_n)}{\lambda_0J_n}\!+\!\sum\limits_{n=1}^N\frac{\lambda_n\varphi_n}{\lambda_0}}
 \end{split}
 }
 \end{equation}

 RSs only serve users in their own coverage, thus the bandwidth utilization of RS$n$ is expressed as
 \begin{equation}\label{equ:Wn}
 \small{
 W_n = \int_0^r\frac{r_0}{r_{nm}(l)}\frac{K_n}{\pi r^2}2\pi l\text dl = K_n\gamma_n
 }
 \end{equation}

 Then $\gamma_n$ is obtained as
 \begin{equation}\label{equ:gamma_n}
 \small{
 \gamma_n\!=\!W_n/K_n\!=\!\frac{2r_0K_n}{r^2}\int_0^r\frac{l\text dl}{r_{nm}(l)}/K_n\!=\!\frac{2r_0}{r^2}\int_0^r\frac{l\text dl}{r_{nm}(l)}
 }
 \end{equation}

 As the resource of each station is shared by active users, the number of users associated with a station evolves like the number of customers in a processor-sharing queue with Poisson arrivals and i.i.d. service times \cite{bonald2003wireless}. As the key property of the processor sharing queue is that the stationary distribution of the number of customers is insensitive to the distribution of service times, the stationary distribution of the number of active users of the BS is $Pr(K0'=k)=(\rho_0)^k(1-\rho_0)$ with mean $E[K_0']=\rho_0/(1-\rho_0)$, where $\rho_0$ is the average traffic load of the BS. Applying Little's law \cite{kleinrock1975queueing}, we get $E[K_0']=\lambda_0'/\mu$. Then the average traffic load of the BS $\rho_0$ can be obtained as
 \begin{equation}\label{equ:rho_0}
 \small{
 \rho_0=\lambda_0'/\lambda_0'+\mu
 }
 \end{equation}

 According to the property of the processor-sharing queue described above, the blocking probability is expressed as
 \begin{equation}\label{equ:P0}
 \small{
 P_{0,blk}=Pr(K_0'\geq W_0^{th}/\gamma_0)=\rho_0^{\lceil W_0^{th}/\gamma_0\rceil}
 }
 \end{equation}
 where $\gamma_0$, $\rho_0$ are calculated by Eq.(\ref{equ:gamma_0}), Eq.(\ref{equ:rho_0}) respectively. Thus the blocking probability of the BS is obtained.

 As RSs just serve users in their own coverage, the actual total number of serving users equals to the arriving users, i.e. $K_n'=K_n$, $\lambda_n'=\lambda_n$. Then the average traffic load of RS$n$ is expressed as
 \begin{equation}\label{equ:rho_n}
 \small{
 \rho_n=\lambda_n/\lambda_n+\mu
 }
 \end{equation}

 Thus, the blocking probability of RS$n$ caused by the resource limitation $\widehat{P}_{n,blk}$ is obtained as
 \begin{equation}\label{equ:Pn}
 \small{
 \widehat{P}_{n,blk}=Pr(K_n\geq W_n^{th}/\gamma_n)=\rho_n^{\lceil W_n^{th}/\gamma_n\rceil}
 }
 \end{equation}
 where $\gamma_n$, $\rho_n$ are calculated by Eq.(\ref{equ:gamma_n}), Eq.(\ref{equ:rho_n}) respectively.

\end{appendix}


\IEEEpeerreviewmaketitle

\end{document}